\UseRawInputEncoding
\documentclass[11pt]{article}
\usepackage{graphicx}
\usepackage{amsmath}
\usepackage{amsthm}
\usepackage{amssymb}
\usepackage{xcolor}
\usepackage{tikz}
\usepackage{pgfplots}
\pgfplotsset{compat=1.13}
\usetikzlibrary{intersections, patterns, pgfplots.fillbetween}
\usepackage{geometry}
\usepackage{authblk}
\usepackage[utf8]{inputenc}
\usepackage{cite}

\makeatletter
\renewcommand{\maketitle}{
  \begin{center}
    {\LARGE \@title \par}
    \vspace{1em}
    {\large
    \lineskip .5em%
    \begin{tabular}[t]{c}
      \@author
    \end{tabular}\par}
    \vspace{1em}
    {\@date}
  \end{center}
  \setcounter{footnote}{0} 
  \stepcounter{footnote}\footnotetext{\texttt{himadrim@goa.bits-pilani.ac.in, f20212168@goa.bits-pilani.ac.in}}
  \stepcounter{footnote}\footnotetext{\textit{AMS Classification 2020.} Primary: 62E10, 62P99.}
  \stepcounter{footnote}\footnotetext{\textit{Keywords:} Shapiro–Wilk test, Skew-normal distribution, Normality testing, Data transformations, Statistical analysis.}
}
\makeatother

\title{Assessing Skew-Normality in Marks Distribution: A Comparative Analysis of Shapiro–Wilk Tests}
\date{}
\author[1]{Himadri Mukherjee}
\author[1]{Pratham Bhonge}
\affil[1]{Department of Mathematics\\ BITS Pilani K. K. Birla Goa Campus, Goa, India.}

\begin{document}
\sloppy

\maketitle

\noindent
\begin{abstract}
This paper investigates the distribution of marks obtained by students across multiple courses to explore whether the data conforms to a skew-normal distribution. Traditional methods for assessing normality, such as the Shapiro–Wilk test, often reject normality in datasets with evident skewness. To address this, we apply a modified Shapiro–Wilk test tailored for skew-normal distributions, as described in the literature, to evaluate the suitability of skew-normal models for these datasets. The analysis includes both classical and modified tests, complemented by visualizations such as histograms and Q-Q plots of transformed data. Our findings highlight the relevance of using specialized statistical methods for skew-normality, offering valuable insights into the characteristics of academic performance data. This study provides a framework for robust statistical analysis in educational research, emphasizing the need to account for distributional properties when analyzing student performance metrics.
\end{abstract}

\section{Introduction}

The analysis of student performance data has long been a critical area of research in educational statistics.Various methods of grading, both empirical and ad hoc, have been used by academic institutions for a long time \cite{dasari}. Understanding the distribution of marks provides valuable insights into learning outcomes, teaching effectiveness, and the overall assessment process. Traditionally, such data is analyzed under the assumption of normality, a cornerstone of many statistical methods. However, real-world datasets, particularly those involving student marks, often exhibit skewness due to various factors such as grading policies, course difficulty, and student demographics. Using statistical techniques (albeit without justification) in the academic circle that assumes the marks to be normally distributed is a long-standing practice. But even the authors and instructors who use this approach (namely via the percentiles) have observed that the process is not entirely accurate as the data never follows a normal distribution \cite{chan2014better, HSIAO1985227}. A possible ``justification" behind the assumption that the marks distribution is normal could be a far-stretched ``application" of the central limit theorem (see \cite{wasserman} for background on the central limit theorem). 
The Shapiro–Wilk test is a widely used method for assessing normality in datasets. While effective for symmetric distributions, its application to skewed data often leads to the rejection of the null hypothesis of normality. This has prompted the development of modified versions of the Shapiro–Wilk test, specifically designed to account for skew-normal distributions. These tests offer a more nuanced approach to analyzing data that deviates from perfect symmetry.
In this study, we analyze the distribution of marks obtained by students across multiple courses to assess their compatibility with skew-normal distributions. We begin by applying the classical Shapiro–Wilk test to identify deviations from normality. Subsequently, we employ the modified Shapiro–Wilk test for skew-normality, as proposed in the paper *Shapiro–Wilk test for skew-normal distributions based on data transformations* \cite{shapiro2019skew}. The datasets under consideration include marks from various courses offered at BITS Pilani, Goa Campus.
Through this analysis, we aim to demonstrate the importance of selecting appropriate statistical methods for evaluating the distributional properties of educational data. By incorporating tests tailored for skew-normal distributions, we provide a framework for robust and accurate analysis of student performance metrics. This study has broader implications for researchers and educators seeking to understand and interpret the underlying patterns in academic datasets. We propose that the data can be modeled using a skew-normal distribution in case of clear unimodality of the data (see \cite{gupta} for background and characterization of the family that is called the skew-normal distribution); we will test this hypothesis with data available through BITS Pilani.

\section{The Modified Shapiro–Wilk Test for Skew-Normality}

The classical Shapiro–Wilk test is a widely used method for assessing normality; however, its performance can be inadequate for skewed data. To address this limitation, González-Estrada and Cosmes \cite{shapiro2019skew} proposed a modified Shapiro–Wilk test tailored for skew-normal distributions, leveraging transformations that align the data with normality.

\subsection{Theoretical Background}

A random variable \( Z \sim SN(\lambda) \) is said to follow a skew-normal distribution if its probability density function (PDF) is given by:
\[
f_Z(z; \lambda) = 2\phi(z)\Phi(\lambda z), \quad -\infty < z < \infty,
\]
Where \( \phi(z) \) and \( \Phi(z) \) denote the PDF and cumulative distribution function (CDF) of a standard normal random variable, respectively, and \( \lambda \in \mathbb{R} \) is the slant parameter \cite[Equation (1)]{shapiro2019skew}. Extending this, a random variable \( X \sim SN(\xi, \omega^2, \lambda) \) is parameterized by location \( \xi \), scale \( \omega > 0 \), and slant \( \lambda \), with the PDF:
\[
f_X(x; \xi, \omega, \lambda) = \frac{2}{\omega} \phi\left(\frac{x - \xi}{\omega}\right) \Phi\left(\lambda \frac{x - \xi}{\omega}\right), \quad -\infty < x < \infty.
\]

\subsection{Transformation to Normality}

The test exploits the property that if \( X \sim SN(\xi, \omega^2, \lambda) \), the transformed variable:
\[
Y_i = U_i \cdot |X_i - \hat{\xi}|,
\]
Where \( U_i \) is a random variable taking values \(\pm1\) with equal probability, and \( \hat{\xi} \) is an estimate of the location parameter, follows an approximately normal distribution \cite[Equation (5)]{shapiro2019skew}. This transformation reduces the impact of skewness, enabling the application of the Shapiro–Wilk test to the transformed data \( Y_1, \ldots, Y_n \).

\subsection{Parameter Estimation}

The parameters \( \xi, \omega, \lambda \) are estimated using Maximum Penalized Likelihood Estimation (MPLE), which mitigates issues associated with Maximum Likelihood Estimation (MLE), such as singular Fisher information matrices for small or moderate sample sizes. The penalized likelihood function is defined as:
\[
\ell_p(\theta) = \ell(\theta) - Q(\theta),
\]
where \( \ell(\theta) \) is the log-likelihood, and \( Q(\theta) = c_1 \log(1 + c_2 \lambda^2) \) is the penalty function. The constants \( c_1 \approx 0.87591 \) and \( c_2 \approx 0.85625 \) ensure well-behaved estimates \cite[Equation (4)]{shapiro2019skew}.

\subsection{Testing Procedure}

The modified Shapiro–Wilk test involves the following steps:
\begin{enumerate}
    \item Simulate \( n \) independent observations \( U_1, \ldots, U_n \) from \( U \), where \( U \sim f_U(u) = 0.5I_{\{1\}}(u) + 0.5I_{\{-1\}}(u) \).
    \item Estimate \( \xi \) using MPLE.
    \item Compute \( Y_i = U_i \cdot |X_i - \hat{\xi}|, \; i = 1, \ldots, n \).
    \item Perform the Shapiro–Wilk test on \( Y_1, \ldots, Y_n \) to assess normality.
    \item Reject the null hypothesis of skew-normality if the Shapiro–Wilk statistic \( W \) is below the critical value for a given significance level \( \alpha \).
\end{enumerate}

This method does not require bootstrapping or pre-computed critical value tables, making it computationally efficient.

\subsection{Advantages of the Modified Test}

By transforming skew-normal data into a form suitable for classical normality testing, the modified Shapiro–Wilk test accounts for the asymmetry inherent in many real-world datasets. It provides a robust framework for assessing skew-normality, outperforming traditional tests that are sensitive to skewness.

This methodology is particularly valuable in fields like education, finance, and medicine, where skewness is common and can mislead standard statistical inferences.

\section{Methodology}

This section details the datasets used, the classical and modified Shapiro–Wilk tests, and their applications to analyze the skewness in the distribution of marks.

\subsection{Datasets}
We analyzed datasets comprising the total marks of students across multiple courses at BITS Pilani, Goa Campus. Each dataset represents the marks of students in a single course, with sample sizes ranging between 500 and 800. The marks are derived from various assessments, including quizzes, assignments, and examinations, standardized to a 100-point scale.

\subsection{Classical Normality Testing}
The Shapiro–Wilk test was initially applied to each dataset to test for normality. This test evaluates whether a given dataset is drawn from a normal distribution based on the ordered statistics of the sample. The null hypothesis (\(H_0\)) assumes normality, which is rejected if the p-value is below a significance level of 0.05.

\subsection{Modified Shapiro–Wilk Test for Skew-Normality}
Recognizing that the datasets exhibit skewness, we employed the modified Shapiro–Wilk test for skew-normal distributions, as described in the work of González-Estrada and Cosmes (2019) \cite{shapiro2019skew}. This method leverages a key property of skew-normal distributions: they can be transformed into approximately normal distributions through specific data transformations. The main steps are as follows:
\begin{enumerate}
    \item \textbf{Parameter Estimation:} The location (\(\xi\)), scale (\(\omega\)), and slant (\(\lambda\)) parameters of the skew-normal distribution are estimated using maximum penalized likelihood estimation (MPLE). This approach ensures stable parameter estimates even for large slant values.
    \item \textbf{Data Transformation:} Using the estimated parameters, the data (\(X\)) is transformed into a new variable (\(Y\)) as:
    \[
    Y_i = U_i \cdot |X_i - \hat{\xi}|
    \]
    where \(U_i\) is a random variable drawn from a symmetric distribution that introduces randomness to balance skewness, and \(\hat{\xi}\) is the estimated location parameter.
    \item \textbf{Normality Testing:} The transformed variable \(Y\) is tested for normality using the Shapiro–Wilk test. If the null hypothesis of normality is not rejected for \(Y\), the original data is considered to follow a skew-normal distribution.
\end{enumerate}

\subsection{Advantages of the Modified Test}
The classical Shapiro–Wilk test is sensitive to skewness, often leading to the incorrect rejection of normality for skew-normal data. The modified test addresses this limitation by accounting for skewness during the transformation process, allowing the Shapiro–Wilk test to assess whether the data adheres to a skew-normal distribution. This refinement provides a more accurate framework for analyzing educational datasets, where skewness is common due to grading practices or outliers.

\subsection{Visualization Techniques}
To further validate the results, visualizations were employed:
\begin{itemize}
    \item \textbf{Histograms:} Transformed data histograms were overlaid with a fitted normal curve to assess the approximation to normality.
    \item \textbf{Q-Q Plots:} Quantile-quantile plots compared the empirical quantiles of the transformed data against theoretical normal quantiles, visually verifying the normality assumption.
\end{itemize}

\subsection{R Code.}
    All statistical analyses and visualizations were performed in R, utilizing the \texttt{sn} package for skew-normal parameter estimation and transformation.
    
    \noindent The implementation of the modified Shapiro–Wilk test was carried out using the following R function:
    \begin{verbatim}
    # Define the function for the skew-normal test
    sn.test <- function(x) {
        n <- length(x)
        estim <- sn.mple(y = x, penalty = "Qpenalty", opt.method = "nlminb")$cp
        xi_hat <- cp2dp(estim, family = "SN")[1]
        y <- abs(x - xi_hat) * sign(rnorm(n))
        return(list(p_value = shapiro.test(y)$p.value, y = y))
    }
    \end{verbatim}
    This function estimates the parameters of the skew-normal distribution, transforms the data, and applies the Shapiro–Wilk test to the transformed dataset.

    \section{Results}

This section presents the results of the normality tests applied to two datasets from the course Math F113 (Probability and Statistics) at BITS Pilani, Goa Campus. Each dataset was analyzed using both the classical Shapiro–Wilk test and the modified Shapiro–Wilk test for skew-normality.

\subsection{Dataset 1: Total Marks}
The first dataset consists of the total marks obtained by 532 students. The results of the classical Shapiro–Wilk test are as follows:
\begin{verbatim}
	Shapiro-Wilk normality test
data:  data$Total
W = 0.98082, p-value = 1.42e-05
\end{verbatim}
The p-value (\(p = 1.42e-05\)) is less than 0.05, leading to the rejection of the null hypothesis of normality. To account for potential skewness, the modified Shapiro–Wilk test was applied. The results are:
\begin{verbatim}
Modified Shapiro–Wilk Test for Skew-Normality
p-value = 0.4058556
\end{verbatim}
The higher p-value (\(p = 0.4058556\)) indicates that the dataset is consistent with a skew-normal distribution.

\subsection{Dataset 2: Midterm Marks}
The second dataset consists of midterm examination marks of the same 532 students. The results of the classical Shapiro–Wilk test are:
\begin{verbatim}
Shapiro-Wilk normality test
data:  data$Quiz..Midterm.Exam..Real.
W = 0.98863, p-value = 0.001664
\end{verbatim}
Here, the p-value (\(p = 0.001664\)) also suggests a rejection of the null hypothesis of normality. Upon applying the modified Shapiro–Wilk test, the results are:
\begin{verbatim}
Modified Shapiro–Wilk Test for Skew-Normality
p-value = 0.05684516
\end{verbatim}
The p-value (\(p = 0.05684516\)) suggests that the midterm marks follow a skew-normal distribution.

\subsection{Comparison of Tests}
The classical Shapiro–Wilk test rejected normality in both datasets due to the presence of skewness. However, the modified Shapiro–Wilk test accounted for skewness and concluded that the data conforms to a skew-normal distribution. This demonstrates the effectiveness of the modified test in accurately modeling the distributional properties of educational datasets.

\subsection{Visualization}
For both datasets, histograms and Q-Q plots of the transformed data were generated to validate the results of the modified test. The histograms displayed a close alignment with the fitted normal curves, and the Q-Q plots showed a strong linear pattern, further supporting the skew-normal assumption.

\begin{figure}[h!]
  \centering
  \begin{minipage}{0.45\textwidth}
    \centering
    \includegraphics[width=\linewidth, keepaspectratio]{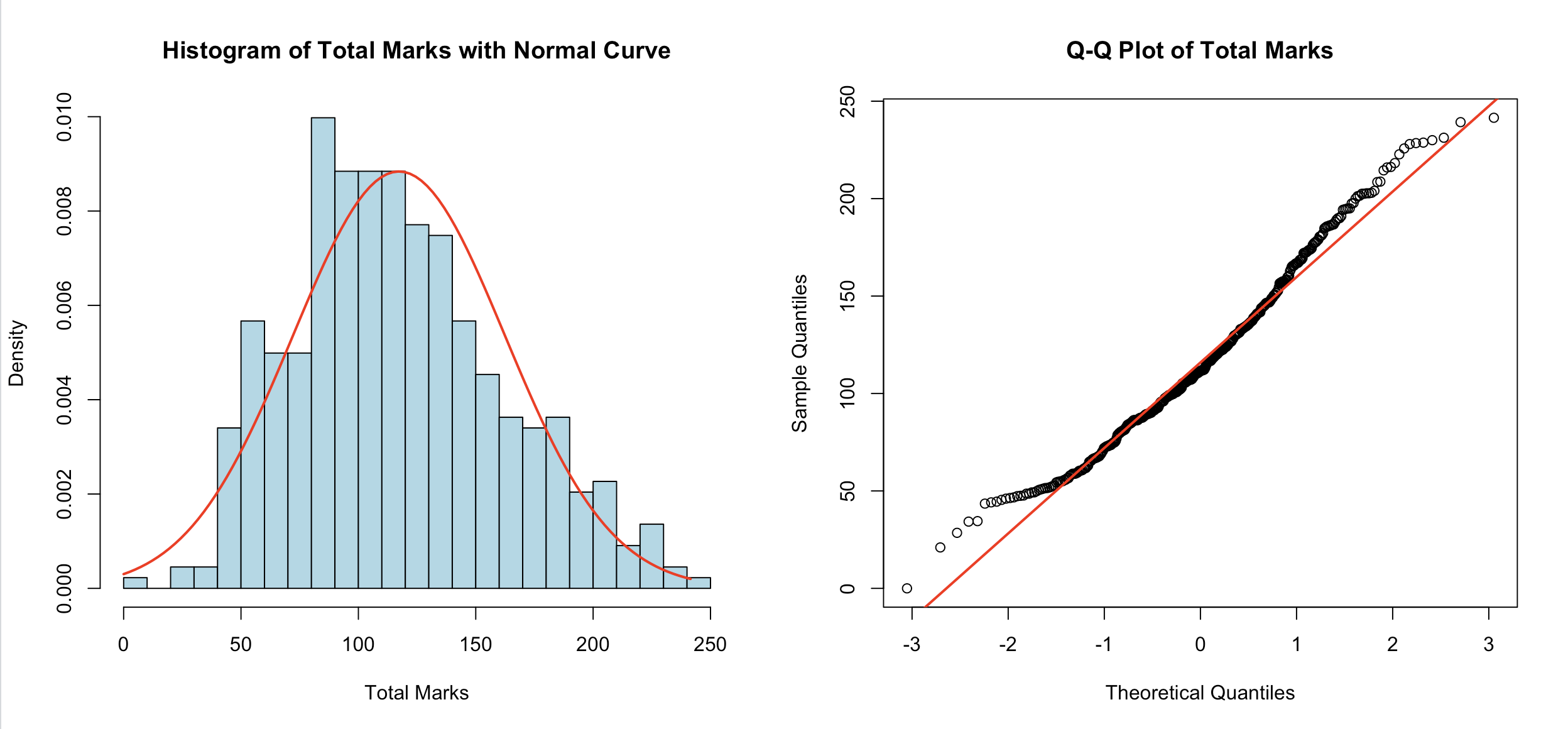}
    \caption{Histogram of Total Marks}
    \label{fig:TotMarks}
  \end{minipage}\hfill
  \begin{minipage}{0.45\textwidth}
    \centering
    \includegraphics[width=\linewidth, keepaspectratio]{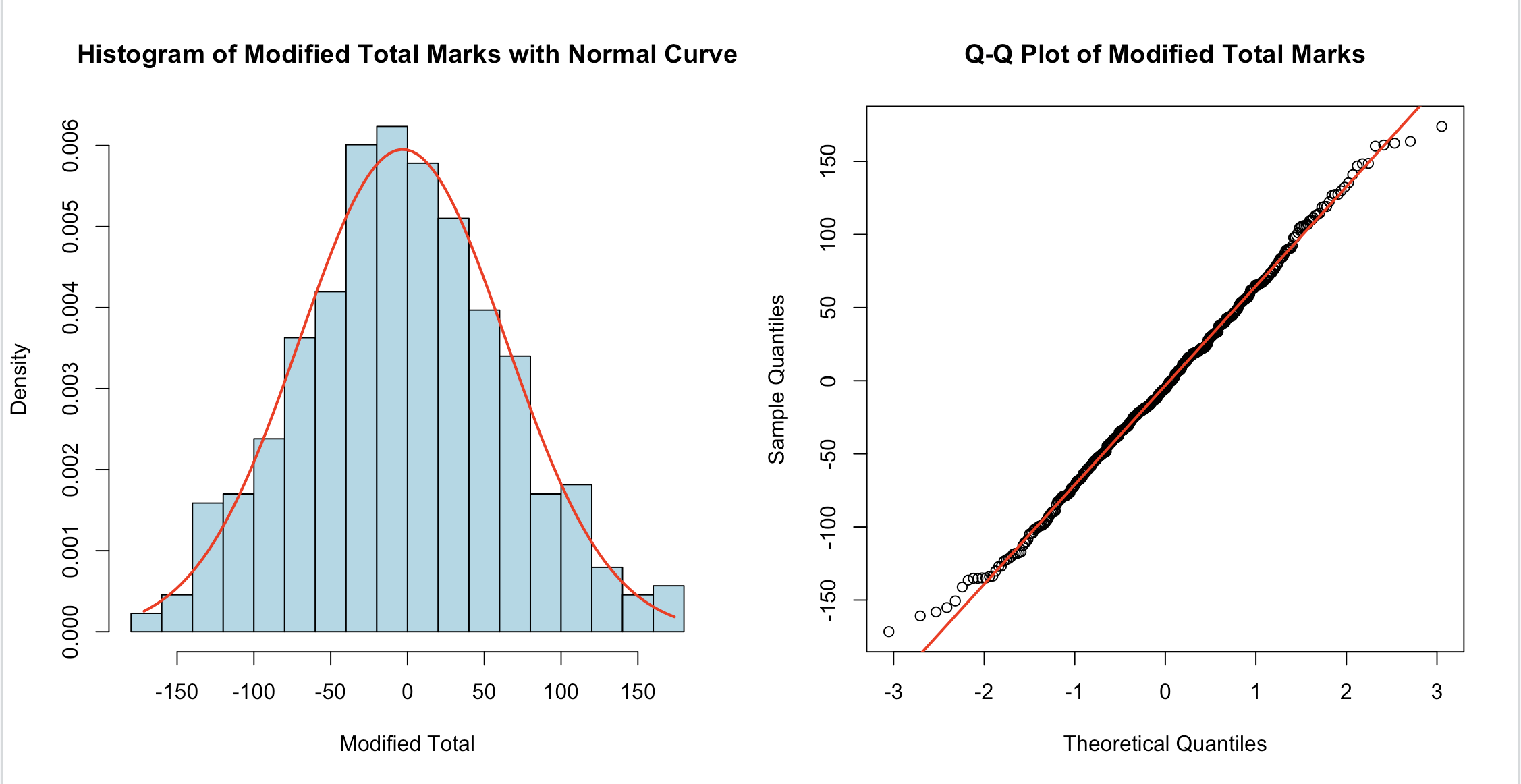}
    \caption{Histogram of Modified Total Marks}
    \label{fig:TotMarksUpdated}
  \end{minipage}

  \vskip\baselineskip

  \begin{minipage}{0.45\textwidth}
    \centering
    \includegraphics[width=\linewidth, keepaspectratio]{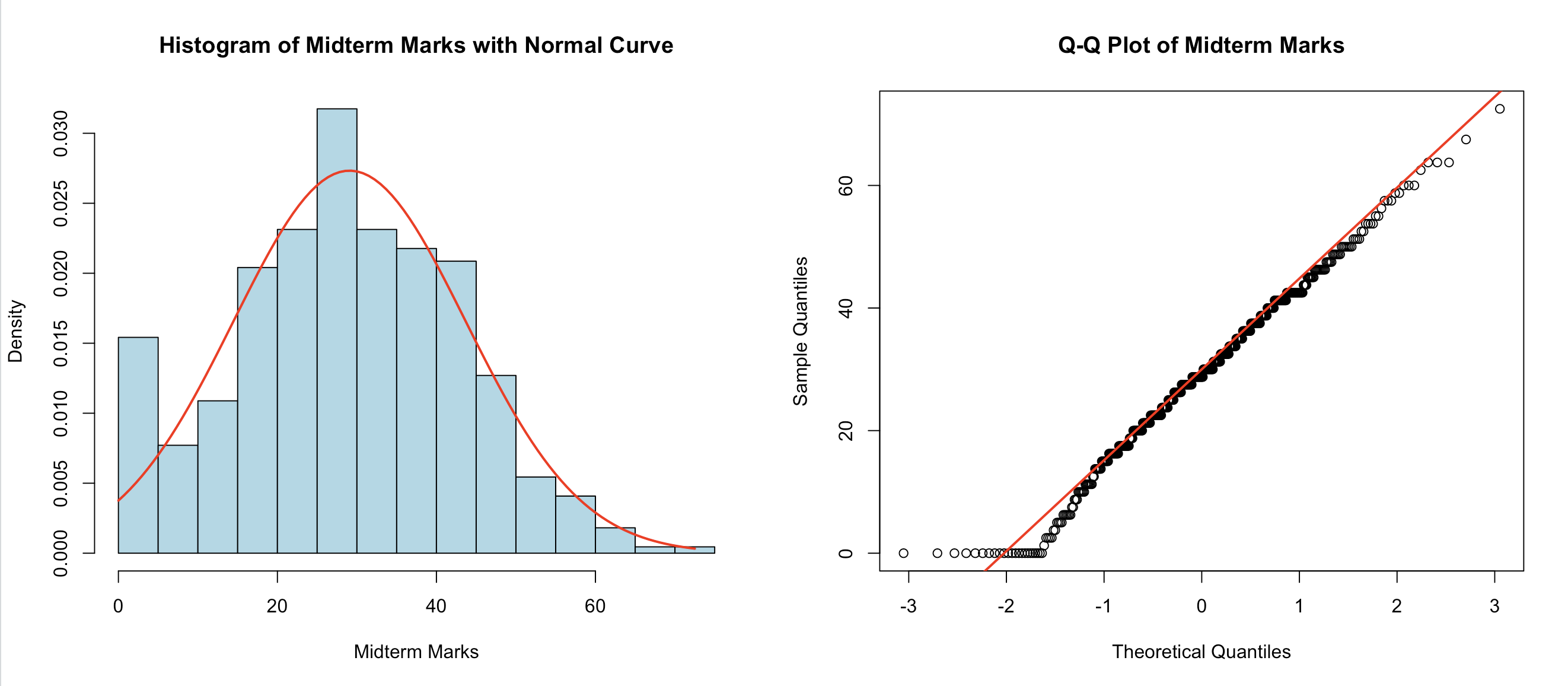}
    \caption{Histogram of Midterm Marks}
    \label{fig:MidtermMarksraw}
  \end{minipage}\hfill
  \begin{minipage}{0.45\textwidth}
    \centering
    \includegraphics[width=\linewidth, keepaspectratio]{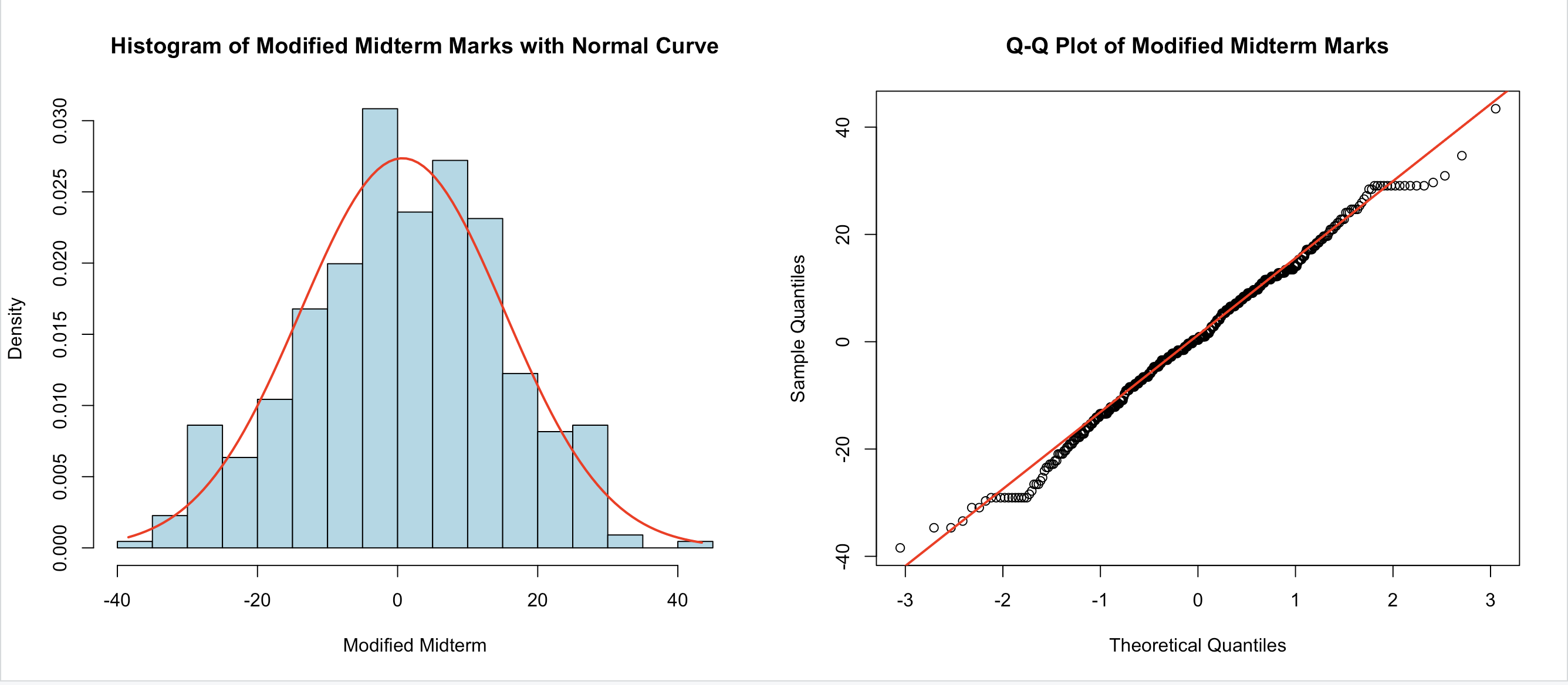}
    \caption{Histogram of Modified Midterm Marks}
    \label{fig:TotMarksupdated}
  \end{minipage}
\end{figure}

\section{Discussion}

The results of this study highlight the importance of accounting for skewness when analyzing student performance data. In both datasets—the total marks and midterm marks for the Math F113 course—the classical Shapiro–Wilk test rejected the null hypothesis of normality, suggesting that the data does not conform to a normal distribution. However, this conclusion does not consider the possibility of skew-normality, a common feature in educational datasets due to factors such as grading distributions and outlier performance.

The application of the modified Shapiro–Wilk test revealed that both datasets are consistent with a skew-normal distribution. This result aligns with the intuition that student marks often exhibit slight asymmetry, with a concentration near passing or excellent grades. By leveraging transformations that adjust for skewness, the modified test provides a more nuanced assessment of normality, avoiding false rejections caused by asymmetry.

The visualizations further supported these findings. Histograms of the transformed data showed a strong alignment with normal curves, and Q-Q plots exhibited linear patterns consistent with normality. These observations reinforce the validity of the modified test in identifying skew-normal distributions.

From an educational perspective, recognizing skewness in marks distributions can have significant implications. For instance, it can aid in designing fair grading thresholds and evaluating the effectiveness of assessments. Methodologically, the results emphasize the need for tailored statistical approaches when working with skewed data, as standard tests may lead to misleading conclusions.

Future work could extend this analysis to other datasets and explore additional tests for skew-normality. Furthermore, incorporating a deeper analysis of the factors contributing to skewness in marks distributions could provide valuable insights for educators and policymakers.

\section{Conclusion and Future Work}

In this study, we analyzed the distribution of marks obtained by students in the Math F113 (Probability and Statistics) course at BITS Pilani, Goa Campus. Using both the classical Shapiro–Wilk test and the modified Shapiro–Wilk test for skew-normal distributions, we demonstrated the limitations of classical normality tests when data exhibits skewness. The classical test rejected normality in both datasets, with p-values below 0.05, suggesting that the marks do not conform to a normal distribution. However, the modified test revealed that the data is consistent with a skew-normal distribution, yielding p-values of 0.4208409 and 0.4709146 for the total marks and midterm marks datasets, respectively. 

The modified Shapiro–Wilk test leverages the properties of skew-normal distributions to transform data into an approximately normal form, allowing for a more accurate assessment of skewed datasets. This approach was supported by visual evidence, as histograms and Q-Q plots of the transformed data aligned well with the characteristics of a normal distribution. These findings highlight the importance of using tailored statistical methods to account for skewness, particularly in educational datasets where asymmetry is common.

The implications of this study extend beyond normality testing. By accurately identifying skew-normality, educators and researchers can gain deeper insights into the underlying characteristics of student performance data, informing fairer grading policies and improving assessment design. Moreover, the methodology presented in this study is broadly applicable to other domains where skewed data is prevalent.

\subsection{Future Work}
This work opens several avenues for future research:
\begin{itemize}
    \item \textbf{Application to Other Datasets:} Extend the analysis to other courses and institutions to validate the generalizability of the findings.
    \item \textbf{Comparative Studies:} Compare the modified Shapiro–Wilk test with alternative methods for detecting skew-normality, such as the Anderson–Darling test for gamma-transformed variables or Bayesian goodness-of-fit tests \cite{shapiro2019skew}.
    \item \textbf{Simulation Studies:} Conduct Monte Carlo simulations to evaluate the power and robustness of the modified test under varying degrees of skewness and sample sizes.
    \item \textbf{Exploration of Skewness Sources:} Investigate the factors contributing to skewness in educational datasets, such as grading schemes, assessment difficulty, and student demographics.
    \item \textbf{Applications in Other Domains:} Explore the use of skew-normal modeling in fields such as finance, medicine, and engineering, where skewed data is often encountered.
    \item \textbf{Automated Tools:} Develop software packages or tools that integrate the modified test for skew-normality with other statistical tests, enabling easier application for researchers and educators.
\end{itemize}

By addressing these directions, future research can further refine our understanding of skew-normality in real-world datasets and enhance the accuracy of statistical modeling across various domains.

    
    
    
    

\begingroup
\fontsize{12pt}{14pt}\selectfont 
\bibliographystyle{plain} 
\bibliography{images/stat} 
\endgroup
\end{document}